\documentstyle[preprint,aps]{revtex}
\begin{document}
\draft
\newfont{\form}{cmss10}
\newcommand{\unity}{1\kern-.65mm \mbox{\form l}}
\newcommand{\k}{\mbox{\form l}\kern-.6mm \mbox{\form K}}
\newcommand{\D}{D \raise0.5mm\hbox{\kern-2.0mm /}}
\newcommand{\A}{A \raise0.5mm\hbox{\kern-1.8mm /}}
\title{Anomalous dimensions and ghost decoupling in a
perturbative approach to the generalized chiral Schwinger model (*)}
\author{A. Bassetto}
\address{Dipartimento di Fisica ``G.Galilei", Via Marzolo 8 -- 35131
Padova, Italy \\ and INFN, Sezione di Padova, Italy}
\author{L. Griguolo}
\address{SISSA, Via Beirut 2 - 34100 Trieste, Italy and INFN, Sezione
di Trieste, Italy}
\maketitle
\begin{abstract}
A generalized chiral Schwinger model is studied by means of perturbative
techniques.
Explicit expressions are obtained, both for bosonic
and fermionic propagators, and compared to the ones derived by means
of functional techniques. In particular a consistent recipe is proposed
to describe the ambiguity occurring in the regularization of the fermionic
determinant.
The role of the gauge fixing term, which is needed to develop
perturbation theory and the behaviour of the spectrum as a function of the
parameters are clarified together with ultraviolet and infrared
properties of the model.
\end{abstract}

(*) This work is carried out in the framework of the European Community
Research Programme ``Gauge theories, applied supersymmetry and quantum
gravity" with a financial contribution under contract SC1-CT92-D789.

\pacs{11.30 Rd}

\narrowtext
\section{Introduction}
\label{prima}
It is well known that two--dimensional gauge theories
admit a consistent interpretation even in presence
of local anomalies: the unitarity is recovered exploiting non--
standard regularization procedure \cite{Sa85} or describing, by
means of an appropriate Wess--Zumino action, the new degrees of
freedom introduced by the anomaly \cite{Ha87}. It is clear that an
analogous higher--dimensional result might be crucial for an
alternative, and perhaps deeper, understanding of the standard
model and of the superstring theory.

Unfortunately at present we have no evidence for a satisfactory
four--dimensional gauge theory with local anomalies: it fails
perturbatively to be either renormalizable or unitary \cite{An91}
while, beyond perturbation theory, there is no real control on it and its
physical interpretation seems indeed very obscure \cite{Mi91}.

These problems do not exist in $d=1+1$, where the non--perturbative region
can be explored by means of powerful techniques like bosonization
\cite{Co75}, conformal field theory \cite{Be84}, form factor approach
\cite{Mu92}, $1/N$ expansion \cite{Ho74}; exact solutions for some classes of
models are also available.

In particular we can study the relation between perturbative and
non--perturbative solution of a simple two--dimensional abelian gauge
model, namely the so--called generalized chiral Schwinger model
\cite{Ba93}. This theory is not completely dull, presenting two quite
different regions in its parameter space: in the first one
the fermionic states are infraparticles described by a Thirring
model while a massive and a massless state appear in the bosonic sector; in
the second case fermions are confined. The interaction gives rise to an
ultraviolet renormalization costant for the fermion field, encoding the
information of ultraviolet scaling.

Previous partial investigations were concerned with the bosonic
sector of the chiral Schwinger model ($r^2 = 1$ in our parameter space),
which is less interesting from the spectrum point of view (no infrared
dressing of the fermions \cite{Ot90} ).

In this paper we firstly present the resummation of the perturbative
expansion for the boson propagator, starting from the Feynman
diagrams: in order to develop the Feynman rules we have to introduce a
gauge fixing.

In the non--perturbative context, where gauge invariance is naturally broken
by the anomaly, this amounts to studying different theories for different
gauge fixings. The limit of vanishing gauge fixing will be performed
after resummation. A lot of interesting features will be hidden in
this limit.

The same propagator will also be obtained by path--integral techniques
(See Appendix). In both procedures we have developed a systematic
method to control the ambiguity related to regularization, clarifying
the way in which the Jackiw--Ramarajan parameter \cite{Sa85} is
produced. Then, studying the bosonic spectrum, we follow the decoupling of
ghost
particles from the theory in the limit of vanishing gauge fixing, to
recover the known result \cite{Ba93}.

The fermionic correlation functions are also examined, leading to
the correct Thirring behaviour in the non--perturbative limit; nevertheless
we find very different ultraviolet scalings before and after the
gauge--fixing removal, related to the appearance of an ultraviolet
renormalization costant.

Decoupling of heavy states is indeed not trivial when anomalies
are present \cite{Fa84}.

\vskip 0.3truecm
\section{Boson propagator and the regularization ambiguity}
\label{seconda}
\vskip 0.3truecm

We want to study the quantum theory in $d= 1+1$ related to the classical
Langrangian density \cite{Ba93}:

\begin{equation}
\label{uno}
{\cal L} = -{1 \over 4} F_{\mu \nu} F^{\mu \nu}+ \bar \psi\,\gamma^{\mu} [i
\partial_{\mu} +
e ({1 +r \gamma_{5} \over 2}) A_{\mu}] \psi,
\end{equation}

$F_{\mu \nu}$ is the usual field tensor, $A_\mu$ the vector potential and
$\psi$ a massless Dirac spinor.
The quantity $r$ is a real parameter interpolating between the vector ($r =
0$)
and the chiral ($r^2 = 1$) Schwinger model. Our notations are

\begin{eqnarray}
\label{due}
g_{00}=- g_{11}&=& 1, \qquad \epsilon^{01}= -\epsilon_{01} =1, \nonumber\\
\gamma^0&=&\sigma_1 ,\qquad \gamma^1=- i\sigma_2, \nonumber\\
\gamma_5&=&\sigma_3, \qquad
\tilde \partial_\mu = \epsilon_{\mu \nu} \partial^\nu,
\end{eqnarray}

$\sigma_i$ being the usual Pauli matrices.

The Green function generating functional is

\begin{equation}
\label{tre}
W [J_\mu; \bar \eta, \eta] = N \int{\cal D} A_\mu {\cal D} \bar \psi
{\cal D} \psi
\,exp\, i \int d^2 x ({\cal L} +{\cal L}_s),
\end{equation}

where $N$ is a normalization constant and

\begin{equation}
\label{quattro}
{\cal L }_s = J_\mu A^\mu + \bar \eta \psi + \bar \psi \eta,
\end{equation}

$J_\mu$, $\eta$ and $\bar \eta$ being vector and spinor sources respectively.

In order to start a perturbative expansion, one has to
break the classical gauge invariance of eq.(\ref{uno}) adding a
gauge-fixing term: we use a generalized Lorentz gauge:

\begin{equation}
\label{cinque}
{\cal L}_{gf} = {1 \over 2 \alpha} (\partial_\mu A^\mu)^2,
\end{equation}

$\alpha \in R$.

In a standard gauge theory physical observables do not
depend on the particular form of the gauge--fixing term. But the
Ward identities of this theory are modified by the presence of
an anomaly in the conservation law of the dynamical current

\begin{equation}
\label{sei}
J^\mu_r (x) = e \bar \psi ({1 - r\gamma_5 \over 2})\gamma^\mu \psi;
\end{equation}

at quantum level gauge invariance is broken and different values of
$\alpha$ do
correspond to different theories. We will be eventually interested in the
limit $\alpha \rightarrow \infty$ (no gauge fixing).

The Feynman propagators associated with ${\cal L} + {\cal L}_{g.f.}$ are given
(in the momentum space) by:

\begin{equation}
\label{sette}
G^{0}_{\mu\nu}(k)=-{i \over k^{2}+i \varepsilon}[g_{\mu \nu}-(1-\alpha)
{k_\mu k_\nu \over k^{2} }],
\end{equation}

\begin{equation}
\label{otto}
S^0 _F  (k) = i{\gamma^{\mu}k_{\mu} \over k^{2} + i \varepsilon},
\end{equation}
and the vertex is
\begin{equation}
\label{nove}
T_\mu = ie ({1 - r\gamma_5 \over 2}) \gamma_\mu.
\end{equation}

Let us look at the perturbative expansion for the boson propagator: it is
well known \cite{Sc62} that, in these kind of theories, the only
non--vanishing one particle--irreducible graph, giving contribution
to the two--point Green functions of $A_\mu$, is:

\begin{equation}
\label{dieci}
\Pi_{\mu\nu}(p)=-\int {d^{2}k \over (2\pi)^{2}}Tr\,[T_\mu S^0 _F
(k)T_\nu S^0 _F  (p-k)].
\end{equation}

The full propagator should be obtained by summing the geometrical series:

\begin{eqnarray}
\label{undici}
G_{\mu\nu}(p)&=&G^{0}_{\mu\nu}(p)+G^{0}_{\mu\rho}(p)\Pi^{\rho\lambda}(p)
G^{0}_{\lambda\nu}(p)+\nonumber\\
&+&G^{0}_{\mu\rho}(p)\Pi^{\rho\lambda}(p)
G^{0}_{\lambda\gamma}(p)\Pi^{\gamma\delta}(p)G^{0}_{\delta\nu}(p)+....
\end{eqnarray}

Actually there is an ambiguity in the calculation of $\Pi_{\mu\nu}$
arising from the need of regularizing the logaritmically divergent
integral in eq.(\ref{dieci}): $\Pi_{\mu\nu}$ does not obey the classical Ward
identity, no matter the regularization we choose, so there is no
privileged choice in fixing the local terms in eq.(\ref{dieci}).
Nevertheless dimensional regularization \cite{Th73} provides a well defined and
systematic way to compute divergent diagrams in absence of $\gamma_5$
couplings.
When $\gamma_5$ occurs, Breitenlohner and Mason (B--M) have developed
in \cite{Br78} a consistent formalism to define $\gamma_5$ as well as the
totally antisymmetric tensor within dimensional regularization; chiral
anomalies appear very naturally in this framework.

In order to reproduce the ambiguity which is intrinsic in the
regularization, we generalize the B--M formalism, showing that there
is a one--parameter family of consistent definitions of $\gamma_5$ and
$\varepsilon_{ \mu \nu}$ in $d=2n$, reproducing the usual one at $d=2$.
The parameter describing the regularization is the origin of the
Jackiw--Ramarajan phenomenon; other schemes leading to analogous
results are presented in \cite{Yu85}, \cite{Ye87} but they are not
obtained as generalizations of the B--M formalism.

We start from the usual properties in $d=2n$

\begin{eqnarray}
\label{dodici}
g_{\mu \nu} g^{\nu}_\lambda &=& g_{\mu \lambda},\qquad
g_{\mu \nu} = g_{\nu\mu},
\nonumber\\
g_{\mu \nu} k^\nu &=& k_\mu,  \qquad g^\mu_\mu = 2n,
\nonumber\\
g_{\mu\nu}\gamma^\nu &=& \gamma_\mu, \qquad \{\gamma_{\mu}, \gamma_\nu\}=
2g_{\mu\nu}\unity.
\end{eqnarray}
As in B--M we write

\begin{equation}
\label{sedici}
g_{\mu \nu} = \bar g_{\mu \nu} + \hat g_{\mu \nu}
\end{equation}

with $\hat g_{\mu \nu}$ carrying indices beyond the ``physical" dimension
$d=2$
and we get:

\begin{eqnarray}
\label{diciasette}
{g_\mu}^\nu \hat g_{\nu \lambda}& =& {\hat g_\mu}^\nu \hat g_{\nu \lambda}
= \hat g_{\mu \lambda}, \nonumber\\
\hat g_{\mu \nu}&=&\hat g_{\nu \mu}, \nonumber\\
\hat g_{\mu \nu} k^\nu &=& \hat k_\mu, \nonumber\\
\bar g_{\mu \nu} \gamma^\nu &=& \bar
\gamma_\mu,\nonumber\\
\hat g_{\mu \nu} \gamma^\nu &=& \hat
\gamma_\mu,
\end{eqnarray}

$\hat \gamma_\mu$ running on the extra dimensions.
Now we just modify the B--M definition of $\epsilon_{\mu \nu}$,
so as to obtain

\begin{equation}
\label{diciannove}
\epsilon _{\mu_{1} \mu_{2}} \epsilon_{\nu_{1} \nu_{2}} = - \Sigma_{\pi
\in S_2}(-1)^{\pi}
\Pi^2_{i =1} (g_{\mu_{i} \nu_{\pi (i)}} - b \hat g_{\mu_{i} \nu_{\pi
(i)}}),
\end{equation}

$S_2$ being the permutation group of two objects ($S_2 = Z_2$) and $b$ a
real parameter; the B--M definition corresponds to $b=1$. It is easy
to prove that:
\begin{equation}
\label{venti}
g_{\mu \nu} \hat \gamma^\nu = \hat g_{\mu \nu} \gamma^\nu = \hat
\gamma_\mu
\end{equation}

\begin{equation}
\label{ventuno}
\{\gamma_\mu, \hat \gamma_\nu \} = \{\hat \gamma_{\mu,} \hat \gamma_\nu\}
= 2 \hat g_{\mu \nu} \unity
\end{equation}

\begin{equation}
\label{ventidue}
\epsilon_{\mu_{1} \mu_{2}} = - \epsilon_{\mu_{2} \mu_{1}}
\end{equation}

\begin{equation}
\label{ventitre}
\hat g^\mu_\mu = 2n -2
\end{equation}

We define $\gamma_5$ as:
\begin{eqnarray}
\label{ventiquattro}
\gamma_5 &=&{1 \over 2 \beta} \epsilon_{\mu \nu }\gamma^\mu \gamma^\nu,
\nonumber\\
\beta^2 &=& 2 n^2 (1-b)^2 + n (1 - 5b) (b-1) + (3b^2 - 2b).
\end{eqnarray}

The normalization is chosen so as to get $\gamma_5^2 = \unity$

This definition coincides with the B--M one ($b=1$) and the limit
$n=1$ ($d=2$) is smooth. Then we define a dual algebra by:

\begin{equation}
\label{venticinque}
\tilde \gamma_\mu = {1 \over 2 \beta} \epsilon_{\mu \nu} \gamma^\nu.
\end{equation}

It follows that

\begin{equation}
\label{ventisei}
\{\tilde \gamma_\mu, \tilde \gamma_\nu\} = 2 \delta_1 \hat g_{\mu \nu}
+ 2 \delta_2 g_{\mu \nu},
\end{equation}

\begin{equation}
\label{ventisette}
\{\tilde \gamma_\mu,  \gamma_\nu\} = {1 \over \beta} \epsilon_{\mu \nu},
\end{equation}

with

\begin{eqnarray}
\label{ventotto}
\delta_1 &=& {b \over 4 \beta^2} [2n - b(2n -2)+b -2], \nonumber\\
\delta_2 &=& -{1 \over 4 \beta^2} [2n - b(2n -2) -1]
\end{eqnarray}

and

\begin{equation}
\label{ventinove}
\gamma_5 = \gamma_\mu \tilde \gamma^\mu.
\end{equation}

{}From eq.(\ref{ventinove}) and the algebras in eqs.(\ref{ventuno}),
(\ref{ventisette}), we are able to find the relevant
anticommutator $\{\gamma_5, \gamma_\mu\}$:
\begin{equation}
\label{trenta}
\{\gamma_5, \gamma_\mu\}=2\gamma_5 \gamma_\mu - 4\tilde \gamma_{\mu}.
\end{equation}

One can easily check that the B--M result is recovered for $b=1$.
For $b \neq 1$ we notice that the anticommutator
$\{\gamma_5, \hat \gamma_\mu\}$ has a term involving also $\bar
\gamma_{\mu}$ and vice versa, at variance with the case $b=1$.
Using eq.(\ref{ventinove}) and the algebra in eqs.(\ref{ventuno}),
(\ref{ventisei}), (\ref{ventisette}), all the traces can be computed.

The parameter $b$ actually
describes a one--parameter family of consistent dimensional
regularizations, which differ by the definition of $\gamma_5$ and
$\epsilon_{\mu \nu}$ and reduce to the ordinary one in physical
dimensions.

The relevant Feynman integral is

\begin{eqnarray}
\label{trentaquattro}
{e^2 \over 16 \pi^2} &\int& d^{2n} k {i \over (p-k)^2 + i \varepsilon}
{i \over k^2 + i \varepsilon} (p-k)^\lambda k^\rho \cdot \nonumber\\
&\cdot&Tr [\gamma_\mu (1+r \gamma_5) \gamma_\lambda \gamma_\nu (1+r \gamma_5)
\gamma_\rho].
\end{eqnarray}

One easily gets:

\begin{eqnarray}
\label{trentacinque}
{e^2 \over 16 \pi^2}&\int & d^{2n} k {i \over (p-k)^2 + i \epsilon}
{i \over k^2 + i \epsilon} (p-k)^\lambda k^\rho (\mu^2)^{1-n}= \nonumber\\
{ie^2 \pi^n \over 16 \pi^2} &\Gamma &^2 (n){\Gamma (2-n) \over \Gamma (2n)}
\Bigl[{g^{\rho \lambda}  \over 2(n-1)} + {p^\rho p^\lambda \over
p^2}\Bigr]
(-{p^2 \over \mu^2})^{n-1}
\end{eqnarray}
$\mu$ being a subtraction mass introduced by dimensional regularization.
The trace part gives :

\begin{eqnarray}
\label{trentasei}
&Tr& [\gamma_\mu \gamma_\rho \gamma_\nu \gamma_\lambda]
\Bigl[ g^{\rho \lambda} {1 \over
2(n-1)} + {p^\rho p^\lambda
\over p^2}\Bigr]=\nonumber\\
& - &4 (g_{\mu \nu} - {p_\mu p_\nu \over p^{2}})+O(n-1),
\end{eqnarray}

\begin{eqnarray}
\label{trentasette}
&Tr&[\gamma_\mu \gamma_5 \gamma_{\lambda} \gamma_\nu \gamma_\rho +
\gamma_\mu \gamma_\lambda \gamma_\nu \gamma_5 \gamma_\rho] [g^{\rho \lambda}
{1 \over 2(n-1)} + {p^\rho p^\lambda \over p^2} ]= \nonumber\\
&-&4 [{\tilde p_\nu
p_\mu \over p^2}+{\tilde p_\mu p_\nu \over p^2}] + O(n-1),
\end{eqnarray}

that do not involve the ``ambiguity" parameter, while

\begin{equation}
\label{trentotto}
Tr [\gamma_\mu \gamma_5 \gamma_\lambda \gamma_\nu \gamma_5
\gamma_\rho] \Bigl[g^{\rho \lambda} {1 \over 2 (n-1)} + {p^\rho p^\lambda
\over p^2}\Bigr]
\end{equation}

consists of an ``unambiguous" piece

\begin{equation}
\label{trentanove}
Tr [\gamma_\mu \gamma_5 \gamma_\lambda \gamma_\nu \gamma_5 \gamma_\rho]
{p^\rho p^\lambda \over p^2} = -2g_{\mu \nu} + 4 {p_\mu p_\nu \over p^2} +
O(n-1)
\end{equation}

and a $b$--dependent one

\begin{eqnarray}
\label{quaranta}
&Tr& [\gamma_\nu \gamma_5 \gamma_\lambda \gamma_\nu \gamma_5
\gamma_\rho] g^{\rho \lambda} {1 \over 2(n-1)} = \nonumber \\
&4&(n-1) [1+2 (1-b^2)]
{1\over 2 (n-1)}
g_{\mu \nu} + O(n-1)= \nonumber\\
&2&[1+2 (1-b^2)] g_{\mu \nu}+ O(n-1).
\end{eqnarray}

Collecting all the terms with the appropriate coefficients
and taking the limit $n=1$, we obtain

\begin{eqnarray}
\label{quarantuno}
\Pi_{\mu\nu} (p) &=& { i e^2 \over 4 \pi} [g_{\mu \nu}
(1 - r^2 (1-b^2)) - (1+r^2) {p_\mu p_\nu \over p^2} + \nonumber\\
&+&r {1 \over p^2} (p_\mu \tilde p_\nu + p_\nu
\tilde p_\mu)].
\end{eqnarray}

The relation between the J--R parameter and $b$ is:

\begin{equation}
\label{quarantadue}
a = r^2(b^2 - 1)
\end{equation}
We notice that a natural bonus of this procedure is to get $a=0$ for $b=1$
(B--M scheme) and for $r=0$ (gauge invariant theory). In the computation we
have disregarded terms with $\hat \gamma_\mu$ and $\hat p_\mu$ on the
external legs: we do not lose any information because the (geometrical) sum
of the vacuum polarization does not involve overlapping divergences.

The resummation is now straightforward and is reported in appendix A.
We define
\begin{equation}
\label{quarantaquattro}
m^2_\pm = e^2 \mu^2_\pm,
\end{equation}
with
\begin{eqnarray}
\label{quarantacinque}
\mu^2_\pm &=&{1\over 8\pi}\Bigl[\alpha (a - r^2)+(1+a)
\pm \nonumber\\
&\pm& \sqrt{[1+a - \alpha (a-r^2)]^2 - 4 \alpha r^2}\Bigr].
\end{eqnarray}
The final result is:

\begin{eqnarray}
\label{cinquantuno}
G_{\mu \nu} (k) &=& i {1 \over p^2 - m^2_+}
{1 \over p^2 - m^2_-} \Bigl[(-p^2 g_{\mu \nu} +(1 - \alpha) p_\mu p_\nu)
+ \nonumber\\
&+&{e^2 \over 4 \pi} (\alpha (a - r^2) g_{\mu\nu} + \alpha (1 + r^2)
{p_\mu p_\nu \over p^2} - \nonumber\\
&-&\alpha r {1 \over p^2} (k_\mu \tilde p_\nu +
p_\nu \tilde p_\mu)\Bigr] ,
\end{eqnarray}
first in the regions $|{m^2_\pm \over p^2}|$ $<1$ which correspond to the
following convergence disc in the complex plane of the coupling constant
$e^2 < {|p^2| \over \mu^2_+}$, and then everywhere by analytic continuation.

We can recover eq.(\ref{cinquantuno}) without resumming the perturbative
series, by exactly computing the generating functional for the bosonic Green
function. This approach is more efficient, expecially in the fermionic case,
where the perturbative expansion is much more involved.

We feel however instructive to obtain the result by summing
Feynman graphs for the bosonic propagator, deferring the
functional integration to the Appendix B.

\vskip 0.5truecm
\section{The bosonic spectrum}
\label{terza}
This section is devoted to the study of the bosonic spectrum of
the theory and of its behaviour in the non--perturbative limit
($|\alpha| \rightarrow \infty$).
We can compare the present result with the exact  non--perturbative
solution obtained in \cite{Ba93}: a non trivial decoupling takes place in
the Hilbert space of the model to recover the spectrum. We can even
understand the emerging of a consistent theory from one which violates
unitarity.

We briefly recall the bosonic content of the non--perturbative
solution: two different regions on the parameters space
($r,a$) admit ``physical" interpretation (no tachyons)

\begin{equation}
\label{cinquantadue}
a > r^2,
\end{equation}
\begin{eqnarray}
\label{cinquantadue2}
 &r^2>1&,\qquad 0<a< r^2 -1,\nonumber\\
 &r^2<1&,\qquad r^2 - 1<a<0.
\end{eqnarray}

In the region described by eq.(\ref{cinquantadue}) a boson of mass

\begin{equation}
\label{cinquantatre}
m^2 = {e^2 \over 4\pi}  {a (a+1-r^2) \over a-r^2}
\end{equation}

exists together with a massless excitation.

In the other region only the massive excitation is physical, the
massless one
being a probability ghost that however can be consistently
expunged from the Hilbert
space by means of a subsidiary condition.

Now the easiest way of reading the physical content in the bosonic sector
of the theories with gauge fixing is to study the singularities
of the propagator eq.(\ref{cinquantuno}): $G_{\mu\nu}$ exhibits
three different poles respectively at $k^2 = m^2_+, k^2 = m^2_- $
and $k^2 =0$. First of all we have to impose the condition

\begin{equation}
\label{cinquantaquattro}
m^2_\pm \geq 0
\end{equation}

which is necessary to have a particle
interpretation for these poles (no tachyons): obviously
inequality (\ref{cinquantaquattro})
selects a particular subregion of the whole parameter space $(\alpha, a, r)$.

It leads to two different sets of inequalities:
\begin{eqnarray}
\label{cinquantacinque}
&\alpha& > 0, \nonumber\\
&a&(a+1 - r^2) >0, \nonumber\\
&1&+a + \alpha (a -r^2) > 0,\nonumber\\
&[&(1 + a)+ \alpha (a - r^2)]-4a \alpha (1+a -r^2)>0
\end{eqnarray}
and
\begin{eqnarray}
\label{cinquantasei}
&\alpha& < 0,\nonumber\\
&a&(a+1-r^2) <0,\nonumber\\
&1&+a + \alpha (a-r^2) <0,\nonumber\\
&[&(1+a) + \alpha (a-r^2) ] - 4a \alpha (1+a-r^2) >0,
\end{eqnarray}
the last inequality being forced from the reality condition
of $m^2_\pm$; we do not consider the limiting situation of vanishing or
equal masses.

It is not too difficult to solve inequalities (\ref{cinquantacinque}) and
(\ref{cinquantasei}) and the allowed regions of the parameters
turn out to be:

${\bf r^2<1}$:
\begin{eqnarray}
\label{cinquantasette}
\alpha > {1\over r^2}\quad &;&\quad a> {1 \over \alpha-1} (1+ \sqrt{\alpha
r^2})^2; \nonumber\\
1< \alpha < {1 \over r^2}\quad &;&\quad 0<a<{1 \over \alpha -1}
(1 - \sqrt{\alpha r^2})^2;\nonumber\\
1< \alpha < {1 \over r^2}\quad &;&\quad
a> {1 \over \alpha -1} (1+ \sqrt{\alpha r^2})^2;\nonumber\\
r^2 < \alpha <1\quad &;&\quad a >0;\nonumber\\
0< \alpha < r^2\quad &;&\quad
{1 \over \alpha -1} (1 - \sqrt{\alpha r^2})^2 < a <r^2 -1;\nonumber\\
0< \alpha < r^2\quad &;&\quad a>0;\nonumber\\
\alpha <0 \quad &;&\quad  r^2 -1 <a <0.
\end{eqnarray}
 ${\bf r^2>1}$:
\begin{eqnarray}
\label{cinquantotto}
\alpha > r^2 \quad &;& \quad a> {1 \over \alpha-1} (1+\sqrt{\alpha
r^2})^2;\nonumber\\
1 <\alpha < r^2 \quad &;& \quad r^2 -1 <a< {1\over \alpha -1}
(1- \sqrt{\alpha r^2})^2;\nonumber\\
1 <\alpha < r^2 \quad &;& \quad
a> {1 \over \alpha-1} (1+ \sqrt{\alpha r^2})^2;\nonumber\\
{1 \over r^2} < \alpha < 1 \quad &;&\quad a> r^2 -1;\nonumber\\
0< \alpha < {1 \over r^2} \quad &;&\quad {1 \over \alpha -1}
(1 - \sqrt{\alpha r^2})^2
<a <0; \nonumber\\
0< \alpha < {1 \over r^2} \quad &;&\quad
a>r^2 -1; \nonumber\\
\alpha < 0 \quad &;& \quad 0<a <r^2 -1.
\end{eqnarray}

For any choice of $r$ and $\alpha$ a particular range of $a$ is free
from tachyons.

The next step is to study the unitarity on these poles by taking the
residues of $G_{\mu\nu} (k)$ at $k^2 = m^2_\pm$ and $k^2 = 0$
and forcing their positivity: we do not give the general result of this
analysis, being the final parameter space rather involved.
Because we are interested in the large $|\alpha|$ behaviour, we give the
details of the unitarity restrictions in the limit
$|\alpha|\rightarrow\infty$. However one can easily verify that for any region
in the parameter space, it never happens that all the
three excitations are ``physical''.

We notice that different regions are selected according to the sign of
$\alpha$.

For $\alpha \rightarrow + \infty$ eqs. (\ref{cinquantasette}) and
(\ref{cinquantotto}) implies:

\begin{equation}
\label{cinquantanove}
a > r^2 + 0({1 \over \sqrt{\alpha}})
\end{equation}

while for $\alpha \rightarrow -\infty$ we get exactly:

\[r^2 -1< a <0 \qquad (r^2 <1)\]
\begin{equation}
\label{sessanta}
0< a<r^2 -1 \qquad (r^2 > 1).
\end{equation}

The masses become, considering the appropriate range in the
two limits:

\begin{equation}
\label{sessantuno}
m^2_+ = {e^2 \over 4\pi} (a-r^2) \alpha + {e^2 \over 4\pi} {r^2 \over
a-r^2} + O({1 \over \alpha}),
\end{equation}

\begin{equation}
\label{sessantadue}
m^2_- = m^2 + O({1 \over \alpha}).
\end{equation}
It is evident that $m^2_+$ goes to infinity with $|\alpha|$, while
$m^2_-$ approaches the generalized J--R mass eq.(\ref{cinquantatre}):
the regions (\ref{cinquantanove}) and (\ref{sessanta}) coincide with
(\ref{cinquanta}), (\ref{cinquantuno}) respectively.

By taking in $G_{\mu\nu}(k)$ the residue at $k^2 = m^2_+$, one gets

\begin{eqnarray}
\label{sessantatre}
&-i&  Res\,\, G_{\mu\nu} (k) |_{k^2 = m^2_+} =
T^+_{\mu\nu} (k) \nonumber\\
&T&^+_{\mu\nu} (k) =
{1 \over e^2/ 4\pi \sqrt{[(1+a)- \alpha(a-r^2)]^{2} -4 \alpha r^2}}\cdot
\nonumber\\
&\cdot&\Bigl[g_{\mu\nu} (-m^2_+ + {e^2 \over 4\pi} \alpha (a -r^2)) +
(1 - \alpha) k_\mu k_\nu +\nonumber\\
& + & \alpha {e^2 \over 4\pi} (1+r^2){k_\mu k_\nu \over m^2_+}
-{e^2 \over 4\pi}4 \alpha r ({\tilde{k}_\mu
k_\nu + \tilde{k}_\nu k_\mu \over m^2_+})\Bigr].
\end{eqnarray}

The determinant of $T^+$  vanishes, so that one eigenvalue is always
zero: this corresponds to the decoupling of the would--be related
excitation. The trace of $T^+$ gives the other eigenvalue:

\begin{eqnarray}
\label{sessantaquattro}
Tr(T^+)& =& {1 \over e^2/4\pi \sqrt{[1+a) - \alpha (a -r^2)]^2 -4 \alpha
r^2}} \cdot \nonumber\\
&\cdot& \Bigl[(k^2_0 + k_1^2) [(1 - \alpha) + \alpha {e^2
\over 4\pi} {(1+r^2)\over m^2_+}] - \nonumber\\
&-&4\alpha r {e^2 \over 4\pi} {k_0 k_1 \over m^2_+}\Bigr]
\end{eqnarray}

that, for $\alpha \rightarrow \pm \infty$ becomes

\begin{equation}
\label{sessantacinque}
Tr(T^+) = -\alpha{(k_0^2 + k^2_1)\over m^2_+} - 4 {r \over a-r^2} {k_0
k_1 \over m^2_+}+ O ({1 \over \alpha})
\end{equation}

In the first region $(\alpha \rightarrow \infty)$ $Tr [T^+]$ is negative
and therefore
the excitation of mass $m_+$ is a probability ghost, while, when
$\alpha\rightarrow -\infty $, it has ``physical" meaning.
We notice that the residue does
not approach a finite value as $m^2_+$ goes to infinity: it does not look
like the naive decoupling one could expect.

The analysis for $m^2_-$ is similar: we define $T^-_{\mu\nu}$ as in
eq.(\ref{sessantaquattro}) and $\det(T^-)$ turns and to be zero.
For large $|\alpha|$:

\begin{eqnarray}
\label{sessantacinque5}
Tr (T^-)&=& {1 \over e^2/ 4\pi} {1 \over a -r^2} [(k^2_0 + k^2_1)
(1- {e^2 \over 4\pi} {(1 + r^2)\over m^2}) -\nonumber\\
&-&4r {e^2 \over 4\pi} {k_0  k_1 \over m^2}]
+ O ({1 \over \alpha}).
\end{eqnarray}

One can easily prove that in both limits

\begin{equation}
\label{sessantasei}
Tr (T^-) > 0.
\end{equation}

The pole at $m^2_-$ is a ``physical" particle and can be identified
with the massive boson of eq.(\ref{cinquantatre}).

We are left with the massless pole at $k^2 =0$: the definition of
$T^0_{\mu\nu}$ implies again a vanishing determinant and

\begin{eqnarray}
\label{sessantasette}
Tr [T^0]&=& {\alpha e^2 /4\pi \over m^2_+ m^2_-} \Bigl[(1 + r^2) (k^2_0 +
k^2_1) - 4r k_0 k_1\Bigr]_{k^2=0}\nonumber\\
&=&{1 \over 4} {1 \over (a-r^2) } (1 \pm r)^2 k^2_0.
\end{eqnarray}

The massless pole appears to be ``physical" in the first range
$(\alpha \rightarrow +\infty, a > r^2)$ and a probability ghost in the
second one: this is exactly the massless particle of \cite{Ba93}.

It is quite unexpected that the residue at the massless pole does
not depend an $\alpha$: the massless sector is totally equivalent
to the one in the non perturbative case. We shall find a similar
behaviour in the fermionic sector.

In conclusion the non--perturbative bosonic spectrum of the
generalized chiral Schwinger model is recovered, starting from the
perturbation theory, in a subtle way. The first window in the
parameter space $(a > r^2)$ corresponds to the situation
$\alpha \rightarrow + \infty$. We obtain the boson of mass $m^2$ from
$m^2_-$ and the massless excitation, together with a ghost of
infinite mass and ``infinite" residue.

The opposite regime $( \alpha \rightarrow -\infty)$ leads to the
window in eq.(\ref{cinquantadue2}) where the massless boson is a ghost and the
infinite massive state has a positive residue. If we perform the limits
$\alpha \rightarrow \pm \infty$, while keeping $k_\mu$ fixed, the
propagator eq.(\ref{cinquanta}) in the two cases, coincides with the
non--perturbative one obtained in \cite{Ba93}: the infinite--mass
boson seems to disappear from the theory if we look at the
bosonic Green function.

But we have seen that its residue grows with $|\alpha|$ and we do not
expect a complete decoupling for more general Green functions (the
fermionic ones for example), as we will see in the next section.

We end by recalling that in the first region unitarity is
obtained by disregarding an infinite--massive ghost (decoupling in
the bosonic Hilbert space), while in the second window no dynamical
mechanism of this type is present and we have to expunge the ghost
excitation by means of a subsidiary condition.

\vskip 0.5truecm
\section{The fermionic spectrum}
\label{quarta}
One of the most interesting feature of the generalized chiral
Schwinger model is the appearence of a dynamically generated massless
Thirring model, describing the fermionic sector of the spectrum in the
first range of the parameters. One can prove \cite{Ba93} that fermionic
correlation functions behave in the infrared limit as the ones of a
massless Thirring model, in the spin--${1 \over 2}$ representation,
with coupling costant

\begin{equation}
\label{sessantotto}
g^2 = {1 - r^2 \over a};
\end{equation}

The fermionic operator solving the quantum equation of motion was
explicity constructed in the form

\begin{equation}
\label{sessantanove}
\psi (x) = exp [F(m^2, x^2)] \psi_T (x)
\end{equation}

with $F(m^2, x^2)$ describing short range bosonic interaction and
$\psi_T (x)$ being the solution of the relevant Thirring theory
\cite{Wt58}.

The ultraviolet limit exhibits a different behaviour, due to the
contribution of the massive boson state: a non--trivial scale dimension
was found, related to an ultraviolet renormalization costant (different
for left and right fermions)

\begin{equation}
\label{settanta}
Z_{L(R)}= ({\Lambda^2 \over m^2})^{-{1\over4}  {(1 \pm r)^2 \over (a -
r^2)}},
\end{equation}

while the c--theorem \cite{Za86} trivially gives the flow between the two
conformally invariant situations (labelled by their central charge c)

\begin{equation}
\label{settantuno}
\Delta c =1.
\end{equation}

For $\alpha \not = \infty$ our solution reproduces only partially
this scenario: as we will see the limit $\alpha \rightarrow \infty$
drastically changes the small distance behaviour of the theory.

In order to study the fermions of the Lagrangian eq.(\ref{uno}) we
compute the two point function

\begin{equation}
\label{settantadue}
S_F (x-y) = < T (\psi (x) \bar \psi (y)) >.
\end{equation}

We recall that local gauge invariance is broken, hence we can extract
meaningful information from the propagator. This Green function can
be computed exactly summing ``by hands" the perturbative expansion or using
its definition in term of $\zeta$--function, namely by an explicit
path--integral
calculation. Obviously both methods give the same result: the functional
integration is very simple, due to the possibility of decoupling the
fermions from the gauge field with a clever change of variables, while the
resummation of the Feynman graphs is rather involved but possible,
following the arguments of \cite{Wu78}. The result is:

\begin{eqnarray}
\label{settantatre}
&S_F& (x)  =  S_L (x) + S_R (x), \nonumber\\
&S_L& (x)  =  Z^L_\alpha S^0_L (x) exp \Bigl \{- i {e^2 (1-r)^2 \over 16 \pi
(m^2_+ - m^2_-)}\cdot \nonumber\\
 &\cdot & \Bigl[(4m^2_-(1-\alpha) + e^2 (1+r)^2 \alpha)
{1 \over m^2_-} \Delta_F (x; m^2_-)- \nonumber\\
& - & (4m^2_+ (1 - \alpha)+e^2(1+r)^2 \alpha) {1 \over m^2_+} \Delta_F
(x; m_+^2)\Bigr]\Bigr\} \nonumber\\
& exp & \Bigl\{{ie^4 (1-r^2)^2 \alpha \over 16 \pi (m^2_+ - m^2)}
\Bigl[{D_F (x; m_-^2) \over m^2_-}-{D_F (x; m^2_+) \over m^2_+} \Bigr]
\Bigr \},
\end{eqnarray}

where

\begin{eqnarray}
\label{settantaquattro}
(i \gamma^\mu \partial_\mu ) ({1 - \gamma_5 \over 2})S^0_L (x)&=
&({1 - \gamma_5 \over
2}) \delta^2 (x), \nonumber\\
\Delta_F (x; m^2)&=& {i \over 2\pi} K_0
(m \sqrt{-x^2 +i\varepsilon}),\nonumber\\
D_F (x; m^2)&=&- {i \over 4\pi} \log(-m^2 x^2 +i\varepsilon),\nonumber\
\end{eqnarray}

and $S_R$ is obtained by changing $r \rightarrow -r$ and $S^\circ_L$
with $S^0_R$.

We notice that the perturbative summation for the fermionic Green function
entails the exchange of bosons with propagator given by
eq.(\ref{cinquantuno}),
which is itself the sum of a geometrical series in the coupling
constant $e^2$. The fermionic Green function requires a convolution
of bosonic propagators in the momentum space; is so doing one needs
a continuation beyond the natural analyticity region $e^2 < {|k^2|
\over \mu^2_+}$. As a consequence we do not expect analyticity of the
fermionic propagator at $e^2 =0$ and indeed eq. (\ref{settantatre})
exhibits a branch point at $e^2 = 0$.
If instead one would compute the fermionic
propagator directly starting from the massless quanta appearing in the
free Lagrangian, one would immediately be confronted with IR singularities
of the perturbative contributions.

We notice that no divergences arise unless $\alpha$ becomes infinite; only a
finite renormalization of the wave function, described by
$Z^{L(R)}_\alpha$, is present

\begin{eqnarray}
\label{settantacinque}
Z^{L(R)}_\alpha & = & ({m^2_+ \over m^2_-})^{\gamma_{L(R)}}, \nonumber\\
\gamma_{L(R)} & = & {(1 \mp r)^2 \over 4} {(1 - \alpha) \over \sqrt{[(1+a) -
\alpha (a-r^2)]^2 - 4 \alpha r^2}}.
\end{eqnarray}

In order to identify the asymptotic states of the theory we study
the large space--like limit of $(x - y)^2$ in eq.(\ref{settantatre}):
the massive propagators do not contribute and we expect $\alpha$--
indipendence in the scaling law, the massless sector being unaware of
the presence of the gauge fixing

\begin{equation}
\label{settantasei}
\lim_{x^2 \rightarrow \infty} S_{L,R} (x) = Z^{L,R}_\alpha
{(m^2_+)^{\rho_1} \over
(m^2_-)^{\rho_2}} (x^2)^{-{1\over 4 }{ (1-r^2)^2 \over a(a+1-r^2)}}
S^0_{L,R} (x),
\end{equation}
where
\[\rho_1={e^4 \over 64\pi^2}{(1-r^2)^2 \over m^2_+ - m^2_-}{\alpha
 \over m^2_+},\]
\[\rho_2={e^4 \over 64\pi^2}{(1-r^2)^2 \over m^2_+ - m^2_-}{\alpha
 \over m^2_-}.\]
The exponent of $x^2$ is actually indipendent of $\alpha$ and coincides
with the one found in the non--perturbative solution.

If we rescale the fermion fields

\begin{eqnarray}
\label{settantasette}
\psi_R & \rightarrow & (Z^R_\alpha)^{-1} \psi_R ,\nonumber\\
\psi_L & \rightarrow & (Z^L_\alpha)^{-1} \psi_L ,\nonumber\
\end{eqnarray}

and we define

\begin{equation}
\label{settantotto}
{(m^2_+)^{\rho_1}\over (m^2_-)^{\rho_2}} = [\mu^2(\alpha)]^{-{1
\over4}{(1-r^2)^2
\over a(a+1-r^2)}},
\end{equation}
we can easily check the dimensional balance in eq.(\ref{settantotto});
the Thirring--like behaviour at large distances is recovered:

\begin{equation}
\label{settantanove}
 \lim_{x^2\rightarrow \infty}S(x) = ( - \mu^2 x^2)^{-{1 \over 4}
{(1-r^2)^2 \over a(a +1-^2)}} S^0 (x).
\end{equation}

The fermionic asymptotic states are the ones found in \cite{Ba93}:
they are constructed with Wick exponentials of the massless field:
no dependence from $\alpha$ can occur.

Let us turn our attention to the opposite regime of the theory,
namely the limit $x^2 \rightarrow 0.$
One finds from eq.(\ref{settantatre})

\begin{equation}
\label{ottanta}
\lim_{ x^2 \rightarrow 0}S_{L,R} (x) =  S^{0}_{L, R}(x).
\end{equation}

Fermions are asimptotically free at variance with the
result \cite{Ba93}, where non trivial scaling was found even
in the ultraviolet regime. It is very easy to check that also the
boson propagator eq.(\ref{cinquantuno}) reduces to the free one eq.(\ref
{sette}) in this situation: we conclude that at small distances the
theory looks like the one of two free Weyl fermions (with a
different normalization forced by our renormalization condition
Eqs.(\ref{settantasette})), carrying central charge $c=1$, and
of a free abelian gauge field carrying vanishing total central charge.
We remark that unless $\alpha \rightarrow \infty$ we are working
with a non--unitary theory and therefore c--theorem does not
hold: no central charge flow exists, the central charge being
1 in the ultraviolet regime as well as in the infrared theory
(massless Thirring model).

We also observe that the limit $e^2 \rightarrow 0$ is possible
in the correlation functions eq.(\ref{settantatre}) as well as in
eq.(\ref{cinquantuno}) and it leads to the ``free" propagators.

The high--energy regime of the present solution is very
different from the non--perturbative one: the recovering
of the unitarity is crucially linked to a change of the
ultraviolet behaviour.

Let us take the limit $\alpha \rightarrow + \infty$ in
eq.(\ref{settantatre})

\begin{eqnarray}
\label{ottantuno}
& \Delta_F &(x;m^2_+) \rightarrow 0 ,\nonumber\\
& \gamma_{L,R} & \rightarrow - {1 \over 4} {(1 \mp r)^2 \over (a -
r^2)},
\end{eqnarray}

\begin{eqnarray}
\label{ottantadue}
\lim_{\alpha \rightarrow + \infty}&S_{L,R}&(x)= S^0_{L,R} (x)
[{e^2 \over 4\pi}
{(a-r^2) \over m^2} \alpha]^{-{1 \over 4} {(1 \mp r)^2 \over
a-r^2}}\nonumber\\
&exp&[- i \pi {(1 \mp r)^2 (a -r^2 \mp r)^2 \over a(a+1 -r^2)(a-r^2)}
\Delta_F (x ,m^2)] \nonumber\\
& exp & [ - {1 \over 4} {(1 -r^2)^2 \over a(a+1 -r^2)} \log (- m^2 x^2 + i
\varepsilon)].
\end{eqnarray}
We get the propagator of \cite{Ba93}
confirming that the non-perturbative solution is recovered.
The large $x^2$ behaviour of eq.(\ref{ottantadue}) is the same of
eq.(\ref{settantatre}): after renormalization of
eq.(\ref{settantasette}), that now is of an infinite type, we get

\begin{equation}
\label{ottantatre}
\mu^2 = m^2.
\end{equation}

The opposite limit is ($\alpha \rightarrow \infty$, $x^2 \rightarrow 0$) gives:

\begin{equation}
\label{ottantaquattro}
S_{L,R} (x) \rightarrow (-m^2 x^2 +i\varepsilon)^{-{1 \over 4} {(1 \mp r)^2
\over a-r^2}} S^0 _{L,R} (x),
\end{equation}

\begin{equation}
\label{ottantacinque}
G_{\mu \nu} (x) \rightarrow  {4 \pi \over e^2} {1 \over (a-r^2)}
\partial_\mu \partial_\nu D_F (x).
\end{equation}

Eqs.(\ref{ottantaquattro}) and (\ref{ottantacinque}) show two important
features: asymptotic
freedom is definitely lost as well as the analyticity of
eq.(\ref{cinquantuno})
in $e^2$ (we notice the appearence of $1/e^2$ terms). We can say that,
sending $\alpha$ to infinity, we shrink to zero the convergence radius
of the power series in $e^2$. One can check that, in this case, the
variation of the central charge
$\Delta c $ from the ultraviolet to the infrared situation is equal to one:
with unitarity c--theorem is recovered.

We can trace the mechanism of the restored unitarity in this way: the original
field $A_\mu$ has no physical degrees of freedom, only a longitudinal
zero norm state made by a physical and a ghost particle (we can check it
in Feynman gauge, for example). This is the original
ultraviolet theory (together with free fermions): the interaction
gives to the $A_\mu$ components $\alpha$ dependent different masses.
As $\alpha \rightarrow + \infty$ the ghost decouples, leaving the
physical particle of mass $m^2$; the long range interaction of
Coulomb--type creates the infrared dressing for the fermions,
leading to a Thirring model.

The drastic change of the dynamical content reflects itself in the
doubling of the U--V central charge and in the divergent character
of the renormalization costant:

\begin{equation}
\label{ottantasei}
Z^{L,R}_\alpha = [{e^2 \over 4\pi} {(a-r^2) \over m^2}  \alpha ]^
{-{1\over 4}
{(1 \mp r)^2 \over a-r^2}}
\end{equation}

with the identification

\begin{equation}
\label{ottantasette}
{e^2 \over 4 \pi} (a - r^2) \alpha = \Lambda
\end{equation}

in eq.(\ref{settanta}). Actually in the limit $\alpha \rightarrow + \infty$
renormalization constant is zero showing that there is no overlap
between the naive perturbative asymptotic states and the effective
solution of the theory.

Looking at expression eq.(\ref{ottantasette}) we can give to $\alpha$ a
different interpretation: we can look at it not as a free parameter in the
perturbative approach but as a cut--off on the non perturbative theory.
One can easily check that

\begin{equation}
\label{ottantotto}
\lim_{\alpha \rightarrow + \infty} (\alpha {\partial \over \partial \alpha})
\log Z_\alpha ^{L, R} = - {1 \over 4} {(1 \mp r)^2 \over (a - r^2)}
\end{equation}
is the ultraviolet scaling, obtained in the usual form of an anomalous
dimension. The regularizing character of $\alpha$ becomes trasparent if
we look at the perturbative expansion of the propagator
eq.(\ref{settantatre}).

Following the suggestions of \cite{Wu78}  we could sum graphs of the type

\begin{eqnarray}
\label{ottantanove}
\Sigma(p)&=&S^{0}_F (p)\int {d^{2}k \over (2\pi)^{2}}
Tr\,\bigl [T_{\mu}G^{\mu\nu}(p-k)T^{\nu}S^{0}_F
(k)\bigr]\cdot\nonumber\\
&\cdot& S^{0}_F (p),
\end{eqnarray}
where $G_{\mu\nu}$
is the propagator eq.(\ref{cinquantuno}). $G_{\mu \nu}$ can be written
as
\begin{equation}
\label{novantuno}
G_{\mu \nu}(k)={1 \over m^2 _+ - m^2 _-}
[G_{\mu \nu}^{+}(m^2 _+;k)-G_{\mu \nu}^{-}(m^2 _- ;k)].
\end{equation}
The contributions of $G_{\mu \nu}^{+}$ and $G_{\mu \nu}^{-}$ are
separately ultraviolet divergent in eq.(\ref{novantuno}), but the divergence
actually cancels in their sum: $\alpha \rightarrow \infty$
corresponds only to the contribution of $G_{\mu \nu}^{-}(m^2;k)$. The
boson $m^2 _+ (\alpha)$ behaves in this scenario as a kind of
Pauli--Villars regulator.

\section{Conclusions}
\label{quinta}

The perturbative solution of the generalized chiral Schwinger
model has been discussed, showing how the spectrum and the ultraviolet
behaviour of Green's functions depend not only on the couplings
($e$ and $r$) and the regularization ambiguity ($a$) but also on a
gauge--fixing
parameter $\alpha$ which is necessary in order to define the free vector
propagator.
However, since gauge symmetry turns out to the broken in the final
solutions owing to the presence of the local anomaly, the introduction
of a gauge fixing term actually amounts to considering inequivalent
theories. The model discussed in \cite{Ba93} corresponds to the limits
$\alpha \rightarrow \pm \infty$ (no gauge fixing).

Accordingly particular regions of the parameter space have been
considered in order to recover the non perturbative solutions in those
limits.
The decoupling of a massive ghost state and the change of the
ultraviolet properties have been discussed when $\alpha \rightarrow +
\infty$: we have observed the transition from an asymptotically free theory
to a theory that exhibits non--trivial scaling behaviour at small
distances. This is related to the non--analyticity in $e^2$ of our result
(after the limit $\alpha \rightarrow + \infty$) and with the
doubling of the ultraviolet central charge.

The appearance of a divergent renormalization costant is intimately
linked to a drastical change of the number of degrees of freedom.
In the infrared regime the massless Thirring model is recovered,
independently of the values of $\alpha$.

\appendix
\section{}
We present the resummation of the perturbative series for the vector
propagator.

At zero order on ${e^2 \over 4\pi}$
we have only the ``free propagator" eq.(\ref{sette}).

At the first order on ${e^2 \over 4 \pi}$ we define the quantity
$A_{\mu\nu}$ as
\begin{equation}
\label{quarantatre}
\Pi_{\mu\nu}(k)= A_{\mu\nu} +i (m^2_+ + m^2_-){1 \over k^2}
(G_{0}^{-1})^{\mu \nu}
\end{equation}
where
\begin{eqnarray}
\label{quarantatretris}
A_{\mu\nu}= &-& i {e^2 \over 4 \pi} [\alpha (a-r^2) (g_{\mu\nu} - {k_\mu
k_\nu \over k^2}) +(1+a) {k_\mu k_\mu \over k^2} - \nonumber\\
&-&r {1 \over k^2}
(\tilde k_\mu k_\nu + \tilde k_\nu k_\mu)],
\end{eqnarray}

\begin{equation}
\label{qurantaquattrobis}
m^2_\pm = e^2 \mu^2_\pm,
\end{equation}
being
\begin{eqnarray}
\label{quarantacinquebis}
\mu^2_\pm &=&{1\over 8\pi}\Bigl[\alpha (a - r^2)+(1+a)
\pm \nonumber\\
&\pm& \sqrt{[1+a - \alpha (a-r^2)]^2 - 4 \alpha r^2}\Bigr].
\end{eqnarray}
We introduce the quantities

\begin{eqnarray}
\label{quarantacinqueter}
B^{\mu\nu}_{\pm}&=& {m^2_\pm \over k^2} G_{0}^{\mu\nu},\nonumber\\
\hat{A}&=&G_{0}AG_{0}.
\end{eqnarray}

It is easy to prove the identity

\begin{equation}
\label{quarantasei}
(\hat{A})_{*}^2= - {(m^2_+ + m^2_-) \over k^2} \hat{A}-{m^2_+ m^2_- \over
k^4}G_{0},
\end{equation}
where we have defined the $*$ product of matrices $\hat{A}$ and $B_{\pm}$ as:
\begin{equation}
\label{extra}
(\hat{A}B)_{*}=\hat{A}*B=\hat{A}G_{0}^{-1}B.
\end{equation}
The equation for $(\hat{A})_{*}^2$ allows to
express higher $A^{\mu\nu}$ insertions as functions of the lowest one:

\begin{equation}
\label{quarantasette}
(\hat{A})_{*}^2 = - \hat{A}*B_+ - \hat{A}*B_- - B_+* B_-
\end{equation}

With the help of eq.(\ref{quarantasette}) we can write the n-th order of
the perturbative expansion as:

\begin{eqnarray}
\label{quarantotto}
(\hat{A}+B_+ + B_-)_{*}^n &=& \sum^n_{m=0} (B_+)_{*}^m * (B_-)_{*}^{n-m} +
\nonumber\\
&+&\sum^{n-1}_{m=0} (B_+)_{*}^m * (B_-)_{*}^{n-1-m} *\hat{A}
\end{eqnarray}

and the complete propagator as

\begin{eqnarray}
\label{quarantanove}
\sum^\infty_{n=0} (\hat{A}&+&B_++B_-)_{*}^n = \sum^\infty_{n=0}
\sum^{n}_{l=0}
(B_+)_{*}^l * (B_-)_{*}^{n-l} + \nonumber\\
&+& \hat{A}*\sum^\infty_{n=0} \sum^{n-1}_{l=0}
(B_+)_{*}^l * (B_-)_{*}^{n-l-1}.
\end{eqnarray}

Taking Lorentz indices into account we get:

\begin{eqnarray}
\label{cinquanta}
G_{\mu\nu} (k)&=&G^0_{\mu\nu} \sum^\infty_{n=0} \sum^{n-1}_{l= 0}
({m^2_+ \over k^2})^l ({m^2_- \over k^2})^{n-l} +\nonumber\\
&+&i{e^2 /4 \pi \over k^4}\Bigl[\alpha (a-r^2)g_{\mu\nu} + \alpha
(1+r^2) {k_\mu k_\nu \over k^2} - \nonumber\\
&-&\alpha r {1 \over k^2}(\tilde k_\mu k_\nu + \tilde k_\nu
k_\mu)\Bigr]\cdot\nonumber\\
&\cdot&\sum^\infty_{n=1}\sum^{n-1}_{l=0} ({m^2_+ \over k^2})^l
({m^2_- \over k^2})^{l-n-1}.
\end{eqnarray}

The series are of geometrical type, and the result is just
eq.(\ref{cinquantuno}).

\section{}

In this appendix we show how to compute the boson propagator eq.(\ref
{cinquantuno}), using a functional method. This kind of calculation is rather
standard in the topologically trivial case, the only subtle point
being the way to implement the Jackiw--Ramarajan ambiguity. The
principal aim of the following discussion is to develop a systematic
formalism to describe the regularization freedom in the functional
approach. Putting to zero the fermionic sources in eq.(\ref{quattro})
we get:

\begin{eqnarray}
\label{auno}
Z [J_\mu]& = &\int {\cal D} A_\mu exp \Bigl[i \int d^2 x
\Bigr(- {1 \over 4} F_{\mu
\nu}F^{\mu \nu} + \nonumber\\
&+&{1 \over 2 \alpha} (\partial_\mu A^\mu)^2 +
J_\mu A^\mu\Bigr)\Bigr]\det [D(A;r)],
\end{eqnarray}

\begin{equation}
\label{adue}
D(A;r)= \gamma^\mu [i \partial_\mu + e ({1 - r \gamma_5 \over 2})A_\mu].
\end{equation}

The functional determinant is obtained by integrating the
fermionic degrees of freedom: $\zeta$--function technique \cite{Hw77}
provides a well defined method to treat determinants
of elliptic operators in any dimension. Really the operator
$D(A;r)$ is of hyperbolic type; so one has to make the computation
in the euclidean space, where the principal symbol \cite{Se67}
is elliptic, and then to continue back the solution to Minkowski. In
two dimensions the calculation can be performed exactly, even in
the non--abelian case \cite{Kd86}: the usual procedure selects a
precise value of the parameter $a$ (as in the B--M scheme): therefore we
propose a generalization of the $\zeta$--function regularization, depending
on a real parameter.

Actually we study a slightly more general problem, considering the
operator

\begin{eqnarray}
\label{atre}
D(A_L, A_R)&=&\gamma^\mu [i \partial_\mu + P_L  A_{R \mu} + P_R A_{L
\mu}], \nonumber\\
P_{R,L}&=&({1 \pm \gamma_5 \over 2}),
\end{eqnarray}
with  $A_{L \mu}$ and $A_{R \mu}$ independent fields.
The identification

\begin{eqnarray}
\label{aquattro}
& e & ({1+r \over 2}) A_\mu = A_{L \mu} \nonumber\\
& e & ({1-r \over 2}) A_\mu = A_{R \mu} \nonumber
\end{eqnarray}
leads immediately to the result eq.(\ref{adue}).

For connections belonging to a trivial $U(1)$ principal--bundle over
the compactified euclidean space we can define

\begin{equation}
\label{acinque}
\hat {\det} [D(A_L, A_R)] = {\det [D(A_L, A_R) \hat D (A_L, A_R)] \over
\det [\hat D (A_L, A_R)]}.
\end{equation}

With $\det$ we mean the standard determinant constructed
by $\zeta$ --function and

\begin{eqnarray}
\label{asei}
\hat D (A_L, A_R)& =& \gamma^\mu \Bigl [i\partial_\mu + P_R (aA_{R\mu} +
cA_{L\mu})+
\nonumber\\ &+& P_L (dA_{R\mu} + bA_{L\mu})\Bigr]
\end{eqnarray}
a,b,c,d, being real parameters. Were the naive factorization property
holding

\begin{equation}
\label{asette}
\det [AB] = \det [A] \det [B],
\end{equation}
no dependence on a,b,c,d would appear in eq.(\ref{acinque}): in this sense
the definition is allowed. We expect that only local terms on the
external fields $A_{\mu L}$ and $A_{\mu R}$ could depend on our
parameters, according to the general claim that different regularizations
cannot modify the non--local part of the effective action.

Using the properties of the complex powers of an elliptic operator
\cite{Se67}, the relevant Seeley-de Witt coefficient \cite{Pb74} and
the Hodge decomposition for $A_{L,R\mu}$, we get (after analytical
continuation to Minkowski space):

\begin{equation}
\label{aottobis}
{\hat {\det} [D (A_L, A_R]  \over \det [i\gamma^{\mu}\partial_{\mu}]}
= exp - {i \over 8\pi}\int d^2x \, {\cal L}_{eff.}(A_R,A_L),
\end{equation}
\begin{eqnarray}
\label{aotto}
&{\cal L}&_{eff.}(A_R,A_L)=
A^\mu _L [g_{\mu, \nu} (1+ a_L) - 2{\partial_\mu
\partial_\nu \over \Box}+ \nonumber\\
&+&({\tilde \partial_\mu \partial_\nu +
\tilde \partial_\nu \partial_\mu \over \Box})]A^\nu_{L}+
A^\mu_R [g_{\mu \nu} (1 + a_R) -\nonumber\\
&-&2 {\partial_\mu
\partial_\nu \over \Box}- ({\tilde \partial_\nu \partial_\mu +
\partial_\nu \tilde \partial_\mu \over \Box})] A^\nu_R + \nonumber\\
&+& 2 A^\mu_L [g_{\mu \nu} (1+b_1) - \epsilon_{\mu \nu}
(1+b_2)] A^\nu_R,
\end{eqnarray}

with

\begin{eqnarray}
\label{anove}
& a_L & = 2b (1 - c), \nonumber\\
& a_R & = 2a (1 - d), \nonumber\\
& b_2 & = ab - (1-c) (1-d) ,\nonumber\\
& b_1 & = -ab - (1-c) (1-d) .\nonumber
\end{eqnarray}

Actually we have only three indipendent parameters because eq.(\ref{anove})
implies

\begin{equation}
\label{adieci}
b^2_2 - b^2_1 = 2 a_L a_R.
\end{equation}

Eq.(\ref{aquattro}) leads to the desired result:

\begin{eqnarray}
\label{aundici}
{\hat {\det} [D(A; r)] \over \det [i\gamma^\mu \partial_\mu]}& = &
exp - {ie^2 \over 8\pi}
\int d^2x A^\mu [g_{\mu \nu} (1 +a) - \nonumber\\
 - (1 + r^2) {\partial_\mu
\partial_\nu \over \Box}+ &r& ({\partial_\mu \tilde \partial_\nu +
\partial_\nu \tilde \partial_\mu \over \Box})] A^\nu,
\end{eqnarray}

\begin{equation}
\label{adodici}
a = {(1+r)^2 \over 2} a_R + {(1-r)2 \over 2} a_L +b_1 {(1-r^2) \over 2}
\end{equation}

The functional $Z[J_\mu]$ turns out to be
(normalized to $\det [i \gamma^\mu \partial_\mu])$:

\begin{equation}
\label{atredici}
Z [J_\mu] = \int {\cal D} A_\mu exp \,i \int d^2 x [- {1 \over 2} A^\mu K_
{\nu \mu} A^\nu + J_\mu A^\mu],
\end{equation}

with
\begin{eqnarray}
\label{aquattordici}
K_{\mu \nu} &=& - g_{\mu \nu}\Box + (1 - \alpha) \partial_\mu
\partial_\nu - {e^2 \over 4 \pi} \Bigl[g_{\mu \nu} (1 +a) - \nonumber\\
&-&(1+r^2){\partial_
\mu \partial_\nu \over \Box} + r ({\partial_\mu \tilde \partial_\nu +
\partial_\nu \tilde \partial_\mu \over \Box}\Bigr].
\end{eqnarray}
The propagator is nothing but the inverse of this operator

\begin{equation}
\label{aquindici}
G_{\mu \nu} (x,y) = i K^{-1}_{\mu \nu} (x, y).
\end{equation}

The inversion of $K_{\mu \nu}$ is performed by means of the Fourier transform
and is very simple although tedious:

\begin{eqnarray}
\label{asedici}
G_{\mu \nu} (x,y)& = &{1 \over m^2_+ - m^2_-} \Bigl[{-e^2 \over 4 \pi}
\alpha (a-r^2) g_{\mu \nu}  - g_{\mu \nu} \Box +\nonumber\\
& +&{1 \over m^2_+}[(1 - \alpha) m^2_+ + \alpha {e^2 \over 4 \pi}
(1+r^2)] \partial_\mu \partial_\nu -\nonumber\\
&-& {e^2 \over 4 \pi} {\alpha r \over m^2_+} (\partial_\mu \tilde
\partial_\nu + \partial_\nu \tilde \partial_\mu)\Bigr] \Delta_F (x,y; m^2_+)+
\nonumber\\
& +& {1 \over m^2_+ - m^2_-}
{e^2 \over 4 \pi}  {\alpha \over m^2_+} \Bigl[ r(\partial_\mu \tilde
\partial_\nu + \partial_\nu \tilde \partial_\mu) - \nonumber\\
&-&(1+r^2) \partial_\mu
\partial_\nu)\Bigr] D_F (x,y) \nonumber\\
&+& m^2_+ \leftrightarrow m^2_- .
\end{eqnarray}
The Fourier transform of $G_{\mu \nu}$ coincides with
eq.(\ref{cinquantuno}).


\begin{thebibliography}{100}
\bibitem{Sa85}{ R.Jackiw and R. Rajaraman, Phys.Rev. Lett. {\bf 54} 1219
(1985),} {H.O. Girotti, H.J. Rothe and K.D. Rothe, Phys. Rev.
{\bf D33} 514 (1986).}
\bibitem{Ha87}{L.D. Faddeev, Phys. Lett. {\bf 145 B}, 81 (1984),}
 {K. Harada and I. Tsutsui, Phys. Lett. {\bf B183} 311
(1987),}
{O. Babelon, F.A. Shaposnik and C.M. Viallet, Phys. Lett. {\bf B177}385
(1986).}
\bibitem{An91} {A. Andrianov, A. Bassetto and R. Soldati, Phys. Rev. {\bf
D44} 2602 (1991).}
\bibitem{Mi91} {P. Mitra, Ann. of Phys. {\bf 211} 158 (1991),}
{F.S. Otto and K.D. Rothe preprint HD-THEP 91 -49 December 1991,}
{F.S. Otto, Phys. Rev. {\bf D47} 623 (1992).}
\bibitem{Co75} {S. Coleman, Phys. Rev. {\bf D11} 2088 (1975)},
{S. Mandelstam, Phys. Rev.{\bf D11} 3026 (1978).}
\bibitem {Be84}A.A. Belavin, A.M. Polyakov and A.B. Zamolodchikov, Nucl. Phys.
{\bf B241} 333 (1984).
\bibitem{Mu92} G. Mussardo, Phys. Rev. {\bf 218} 524 (1992).
\bibitem{Ho74} G. `T Hooft, Nucl. Phys. {\bf B75} 461 (1974).
\bibitem{Ba93} A. Bassetto, L. Griguolo and P. Zanca, Padova preprint
DFPD 93/TH/34 (1993), Phys. Rev. {\bf D} to appear.
\bibitem{Ot90} {H.J.Rothe and K.D. Rothe, Phys. Rev. {\bf D40} 545,
(1989)}
{F.S. Otto, A. Recknagel and H.J.Rothe, Phys. Rev. {\bf D42} 1203
(1990).}
\bibitem{Fa84} E.Farhi and E.D'Hooker, Nucl. Phys. {\bf B248} 59 (1984).
\bibitem{Sc62} Y.Frishman, Proc. Mexican Summer Institute, Lecture Notes
in Physics, {\bf Vol 32} 118 (Springer, Berlin 1973).
\bibitem{Th73} G.'t Hooft and M.Veltman, Diagrammar CERN 73-9 (1973)
\bibitem{Br78} P.Breitenlohner and D.Maison, Comm. Math. Phys.
{\bf 52} 11 (1978).
\bibitem{Yu85} G.T.Thompson and H.L.Yu, Phys. Lett. {\bf 151B} 119
(1985).
\bibitem{Ye87} H.L.Yu and W.B.Yeung, Phys. Rev. {\bf D35} 3955 (1987).
\bibitem{Wt58} W.Thirring, Ann. of Phys. {\bf 3} 91 (1958).
\bibitem{Za86} {A.B.Zamolodchikov, JEPT Lett. {\bf 43} 730 (1986),}
{S.J. Nucl. Phys. {\bf 46} 1090 (1987).}
\bibitem{Wu78} I.O.Stamatescu and T.T.Wu, Nucl. Phys. {\bf B143} 503
(1978).
\bibitem{Hw77} S.W.Hawking, Comm. Math. Phys. {\bf 55} 133 (1977).
\bibitem{Se67} R.T.Seeley, Am. Math. Soc. Proc. Symp. Pure Math. {\bf 10}
288 (1967).
\bibitem{Pb74} P.B.Gilkey: The index theorem and the heat--equation
Boston Publish of Perish (1974)
\bibitem{Kd86} K.D.Rothe, Nucl. Phys. {\bf B269} 269 (1986).
\end{thebibliography}
\end{document}